\documentclass[apjl]{emulateapj}
\usepackage{verbatim}
\usepackage{natbib}
\usepackage{apjfonts}
\usepackage{hyperref}

\def\lae{\mathrel{<\kern-1.0em\lower0.9ex\hbox{$\sim$}}}
\def\gae{\mathrel{>\kern-1.0em\lower0.9ex\hbox{$\sim$}}}

\def\mone{$^{-1}$}
\def\mtwo{$^{-2}$}

\newcommand{\Msol}{$M_{\odot}$}

\begin{document}

\shortauthors{TREMBLAY ET AL.}
\shorttitle{``BEADS ON A STRING'' STAR FORMATION BETWEEN TWO MERGING EARLY TYPE GALAXIES}

\title{a thirty kiloparsec chain of ``beads on a string'' star formation 
\\ between two merging early type galaxies in the core of a strong-lensing galaxy cluster}

\author{Grant R.~Tremblay\altaffilmark{1}}      
\author{Michael D.~Gladders\altaffilmark{2}}
\author{Stefi A.~Baum\altaffilmark{3}}
\author{Christopher P.~O'Dea\altaffilmark{3}}
\author{Matthew B.~Bayliss\altaffilmark{4,5}}
\author{~~~~~~~~~~~~~~~~~~~~~~~~~~~~~~Kevin C.~Cooke\altaffilmark{3}}
\author{H\r{a}kon Dahle\altaffilmark{6}}
\author{Timothy A.~Davis\altaffilmark{1}}
\author{Michael Florian\altaffilmark{2}}
\author{Jane R.~Rigby\altaffilmark{7}}
\author{~~~~~~~~~~~~~~~~~~~~~~~~~~~~~~~~~~~~~~~~~~~~~~~~~~~~~~~~~~~~~~~~~~~~~~~~~~~~~~~~~~~~~~~~~~~~~~~~~~~~~~~~~~~~~~~~~~Keren Sharon\altaffilmark{8}}
\author{Emmaris Soto\altaffilmark{9}}
\author{Eva Wuyts\altaffilmark{10}}

\altaffiltext{1}{European Southern Observatory, 
Karl-Schwarzschild-Str.~2, 85748 Garching bei M\"{u}nchen, Germany; grant.tremblay@eso.org}

\altaffiltext{2}{Department of Astronomy \& Astrophysics and Kavli Institute for Cosmological Physics, University of Chicago, 5640 S.~Ellis Ave., Chicago, IL 60637, USA}

\altaffiltext{3}{Chester F.~Carlson Center for Imaging Science and School of Physics and Astronomy, Rochester Institute
of Technology, 84 Lomb Memorial Drive, Rochester, NY 14623, USA}

\altaffiltext{4}{Department of Physics, Harvard University, 17 Oxford St., Cambridge, MA 02138, USA}

\altaffiltext{5}{Harvard-Smithsonian Center for Astrophysics, 60 Garden Street, Cambridge, MA 02138, USA}

\altaffiltext{6}{Institute of Theoretical Astrophysics, University of Oslo, P.O. Box 1029, Blindern, N-0315 Oslo, Norway}

\altaffiltext{7}{Observational Cosmology Laboratory, NASA Goddard Space Flight Center, Code 665, Greenbelt, MD 20771 USA}

\altaffiltext{8}{Department of Astronomy, University of Michigan, 500 Church Street, Ann Arbor, MI 48109, USA}

\altaffiltext{9}{Department of Physics, The Catholic University of America, 200 Hannan Hall, Washington, DC 20064, USA}

\altaffiltext{10}{Max-Planck-Institut f\"{u}r Extraterrestrische Physik, Postfach 1312, Giessenbachstr., 85741 Garching bei M\"{u}nchen, Germany}

\slugcomment{Accepted for Publication in ApJ Letters, 24 June 2014}

\begin{abstract} 
New {\it  Hubble Space Telescope} ({\it HST})  ultraviolet and optical
imaging  of  the  strong-lensing  galaxy  cluster  SDSS  J1531+3414
($z=0.335$)   reveals  two centrally dominant
elliptical  galaxies participating  in  an ongoing  major merger.  The
interaction is at least somewhat  rich in cool gas, as the merger is associated with
a  complex network  of nineteen  massive 
superclusters  of  young stars (or small tidal dwarf galaxies)
separated by $\sim 1$ kpc in projection from one another, combining to
an estimated total star formation rate of $\sim 5$ \Msol\ yr\mone. The
resolved young stellar superclusters   
are threaded by narrow  H$\alpha$, [O{\sc~ii}], and blue excess filaments
arranged in  a network spanning $\sim  27$ kpc across the two merging
galaxies.
This morphology is strongly reminiscent of the well-known ``beads on a
string'' mode of star formation  observed on kpc-scales in the arms of
spiral  galaxies, resonance  rings,  and in  tidal  tails between interacting galaxies. Nevertheless, the arrangement  of this star formation
relative to the  nuclei of the two  galaxies is difficult to interpret in a dynamical sense, as no known
``beads  on  a  string''   systems  associated  with  kpc-scale  tidal
interactions  exhibit such  lopsided  morphology relative to the merger participants.  In  this Letter  we
present the images and follow-up spectroscopy, and 
discuss possible physical interpretations for the unique arrangement of the young stellar clusters. 
While we suggest that this morphology is likely to be dynamically short-lived, a more quantitative understanding awaits necessary 
multiwavelength follow-up, including optical  integral field spectroscopy, ALMA sub-mm interferometry, and  {\it Chandra}  X-ray
imaging. 
\end{abstract}

\keywords{
galaxies: clusters: general ---
gravitational lensing: strong ---
galaxies: interactions ---
galaxies: star formation ---
galaxies: clusters: individual (SDSS J1531+3414)}

\section{Introduction}

SDSS J1531+3414 is a strong lensing cluster of galaxies at redshift
$z=0.335$ (e.g., \citealt{hennawi08,oguri09,oguri12,gralla11,bayliss11,postman12}). Ground-based imaging of the cluster
center reveals high surface brightness gravitational arcs from at least two 
lensed background galaxies at $z\approx1.1$ and $z\approx1.3$
\citep{hennawi08,bayliss11}, enabling weak- and strong-lensing analysis
that yields a cluster mass of 
$M_{200}=5.13^{+1.33}_{-1.19} \times 10^{14}~h^{-1}$ \Msol\ 
\citep{oguri12}. The cluster core was recently imaged as part of
strong lensing imaging program with the {\it Hubble Space Telescope}
({\it HST}).  These new high spatial resolution images reveal that the
two giant elliptical galaxies in the cluster
core are likely in the process of merging. Most remarkably, the merger is
associated with ongoing or very recent star formation arranged in a
$\sim 27$ kpc-scale network of young stellar superclusters separated
$\sim 1$ kpc in projection from one another along faint and narrow
filaments, resembling a broken pearl necklace.  While indeed strongly
reminiscent of the well-known ``beads on a string'' mode of star
formation frequently observed in the arms of spiral galaxies,
resonance rings, and tidal arms that bridge interacting galaxies
(e.g., \citealt{elmegreen96}), the morphology and orientation of this
particular chain is unlike any known merging system, and the
phenomenon is rarely (if ever) observed in giant early type galaxies
(e.g., \citealt{kaviraj12}).

In this Letter we present the new {\it HST} images of the merging
elliptical galaxies and the associated network of young stellar
superclusters.  Details of the observations are given in Table~\ref{tab:tab1} and \S2, and the
results presented in \S3 are discussed in \S4.  Throughout this Letter
we assume $H_0 = 71$ km s$^{-1}$ Mpc$^{-1}$, $\Omega_M = 0.27$, and
$\Omega_{\Lambda} = 0.73$.  In this cosmology, 1\arcsec\ corresponds
to 4.773 kpc at the redshift of the two merging ellipticals in the
cluster center ($z=0.335$), where the associated luminosity and
angular size distances are 984.4 and 1756.1 Mpc, respectively, and the
age of the Universe is 9.954 Gyr.

\begin{deluxetable*}{ccccccccc}
\tabletypesize{\footnotesize}  
\tablecaption{summary of observations}
  \tablehead{
    \colhead{Observatory} &
    \colhead{Instrument} &
    \colhead{Filter / Config.} &
    \colhead{Waveband / Central $\lambda$ / Line} &
    \colhead{Integration Time} &
    \colhead{Obs. Date} &
    \colhead{Comment} 
}
  \startdata
{\it HST}  & WFC3 / UVIS &  F390W  & Rest Frame NUV / 3923 \AA &  2256 sec & 6 May 2013 & Young stellar component   \\
\nodata    &  \nodata &  F606W  & Blue optical / 5887 \AA &  1440 sec & \nodata & Includes [O~{\sc ii}], H$\beta$   \\
\nodata    &  \nodata &  F814W  & Optical / 8026 \AA &  1964 sec & \nodata & Includes H$\alpha$+[N~{\sc ii}]   \\
\nodata    &  WFC3 / IR &  F160W  & Red optical / 1.537 $\mu$m &  912 sec & \nodata & Old stellar component   \\
\cutinhead{{\sc follow-up observations}}
NOT & ALFOSC  &  Grism \#7 / 1\arcsec   &  5260 \AA\ /    [O~{\sc ii}], Ca~{\sc ii~H\&K}  &  2400 sec  & 29 Apr 2014                       & Redshift confirmation \\
\nodata & \nodata  &  Grism \#5 / 2\farcs5   &  7000 \AA\ / H$\alpha$+[N~{\sc ii}]    & \nodata   & \nodata                       & \nodata \\
IRAM 30-meter    &  EMIR &  \nodata & 86 GHz / Rest frame CO$(1-0)$ &  2880 sec & 22 Dec 2013 & Non-detection 
  \enddata
\label{tab:tab1}
\end{deluxetable*}

\section{Observations}
Initially intended to study the higher redshift background galaxies
gravitationally lensed by the cluster, {\it HST} imaging of
SDSS J1531+3414 was obtained in Cycle 20 (GO
program 13003, PI: Gladders) with the Wide Field Camera 3 (WFC3,
\citealt{dressel12}). The target was observed with four broadband filters  over three {\it
  HST} orbits, resulting in broad and essentially gap-free coverage
over the near-ultraviolet (NUV) and optical wavelength range.
This Letter also presents follow-up Nordic Optical Telescope (NOT) 
slit spectroscopy and a short IRAM 30-meter CO$(1-0)$ observation. More information 
regarding all new data presented in this paper can be found in Table \ref{tab:tab1}.

The {\it HST} products were calibrated and reduced using the on-the-fly
recalibration pipeline and the {\sc Astrodrizzle} / {\sc DrizzlePac}
packages provided by the Space Telescope Science Institute \citep{fruchter10}.
Images were drizzled to the same 0\farcs03 pixel frame using a 0.8
``\texttt{pixfrac}'' Gaussian kernel.   An RGB ``pretty
picture'' composite combining all four filters was made using the {\sc
  Trilogy} code\footnote{\url{http://www.stsci.edu/~dcoe/trilogy/}}
kindly provided by Dan Coe (see \S2 of \citealt{coe12}) and the CLASH
team \citep{postman12}.  
More details pertaining to reduction of the {\it HST} data can be found in \citet{bayliss13}.

\begin{figure*}
\begin{center}
\includegraphics[scale=0.5]{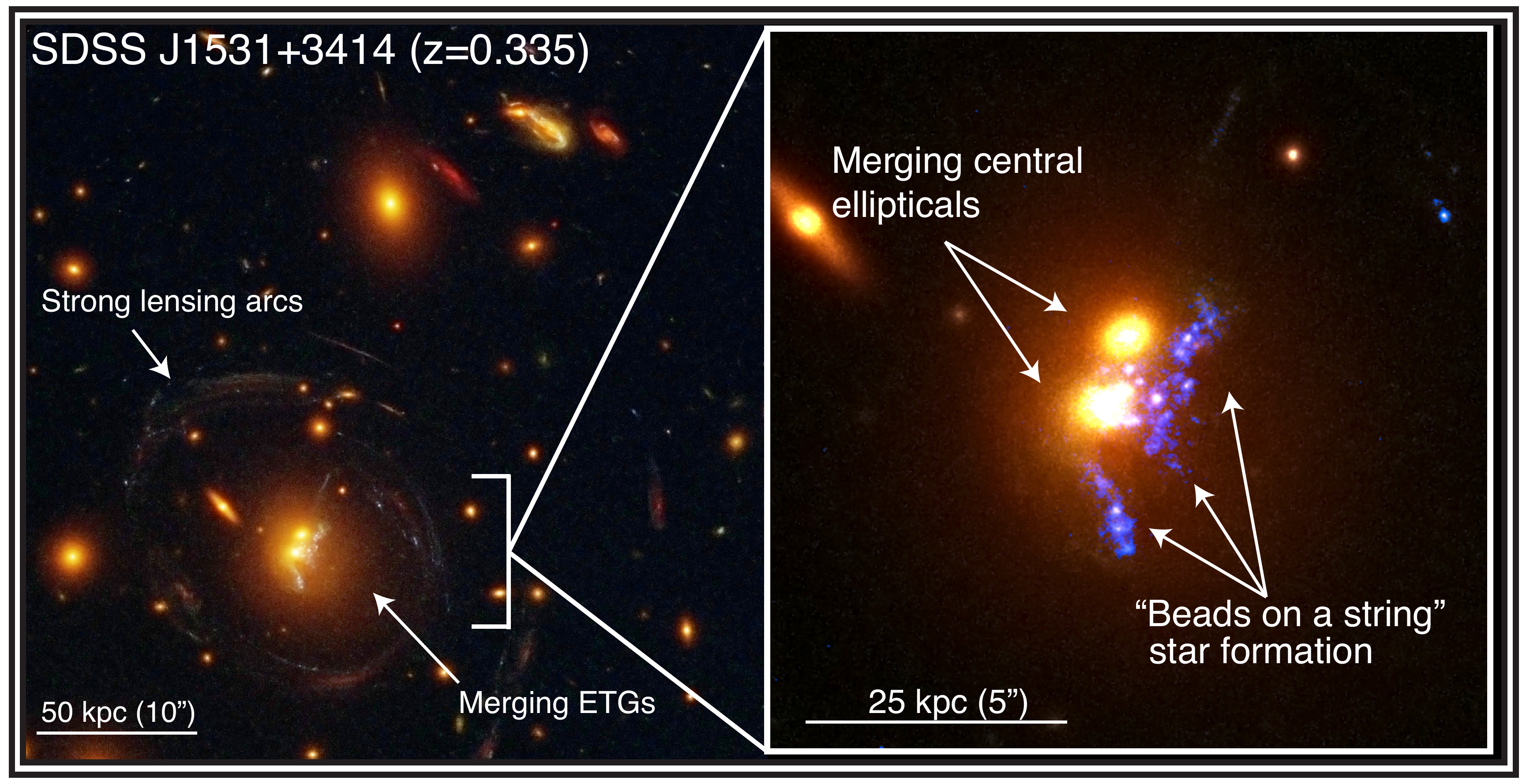}
\end{center}
\vspace*{-3mm}
\caption{A four-color  composite {\it HST}/WFC3 image  of the strong
  lensing  cluster  SDSS  J1531+3414  and its  two  central  brightest
  cluster  galaxies that are likely undergoing a major merger.  The
  F160W and F814W  images are shown in yellow/orange,  the F606W image
  is  shown  in  green,  and  the F390W  image  containing  rest-frame
  near-ultraviolet emission  from young stars is assigned  to the blue
  channel.  ({\it left}) A wide ($\sim 200\times200$ kpc$^2$) view of the
  galaxy  cluster. Tangential gravitational  arcs from  strongly lensed
  background galaxies  are clearly  seen. ({\it right})  A zoom-in  on the
  left-hand   panel,   showing  the   two   merging  central   cluster
  galaxies. Bright NUV emission associated with ongoing star formation
  is observed in blue. }
\label{fig:fig1}
\end{figure*}

\begin{figure*}
\begin{center}
\includegraphics[scale=0.45]{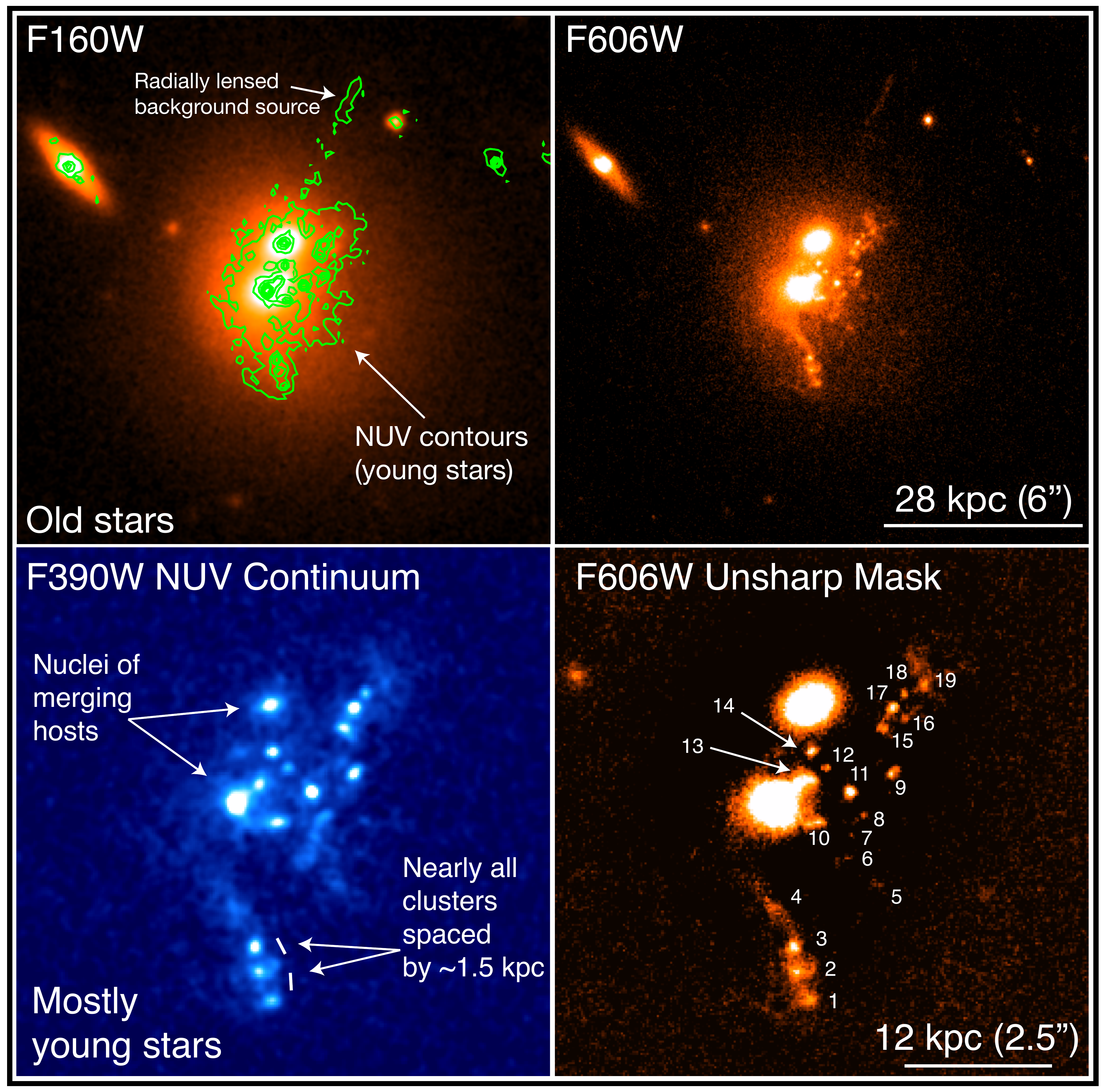}
\end{center}
\vspace*{-3mm}
\caption{ ({\it top left}) F160W image of the old stellar population associated
with the two merging elliptical galaxies. NUV continuum emission (from the F390W 
image) is overlaid in green contours, marking the location of the young 
stars relative to the nuclei and old stellar envelopes of the merging
host galaxies. The linear NUV feature marked
with the arrow in this panel is a radial arc from a lensed
background galaxy (Sharon et al.~2014, in prep). 
({\it top right}) F606W image of both the old and young stellar component. 
({\it bottom left}) 
  F390W rest-frame NUV continuum image (zoomed-in slightly with respect
to the top two panels). The image has been lightly smoothed
  with a  gaussian kernel. The  NUV emission arises mostly  from young
  stars associated  with ongoing or recent star  formation. The nuclei
  of the two  merging elliptical host galaxies are  labeled with white
  arrows. ({\it bottom right})
  Unsharp mask of the F606W image revealing residual fluctuations from
  the overall surface  brightness profile, possibly due to [O~{\sc ii}] and blue excess emission. The stellar superclusters to which we directly refer in the text
are numerically labeled in this panel. The centroids of all panels are aligned, with East left 
and North up. The top row of images shows a slightly wider field of view than 
the bottom row.  
  }
\label{fig:fig2}
\end{figure*}

\begin{figure*}
\begin{center}
\includegraphics[scale=0.48]{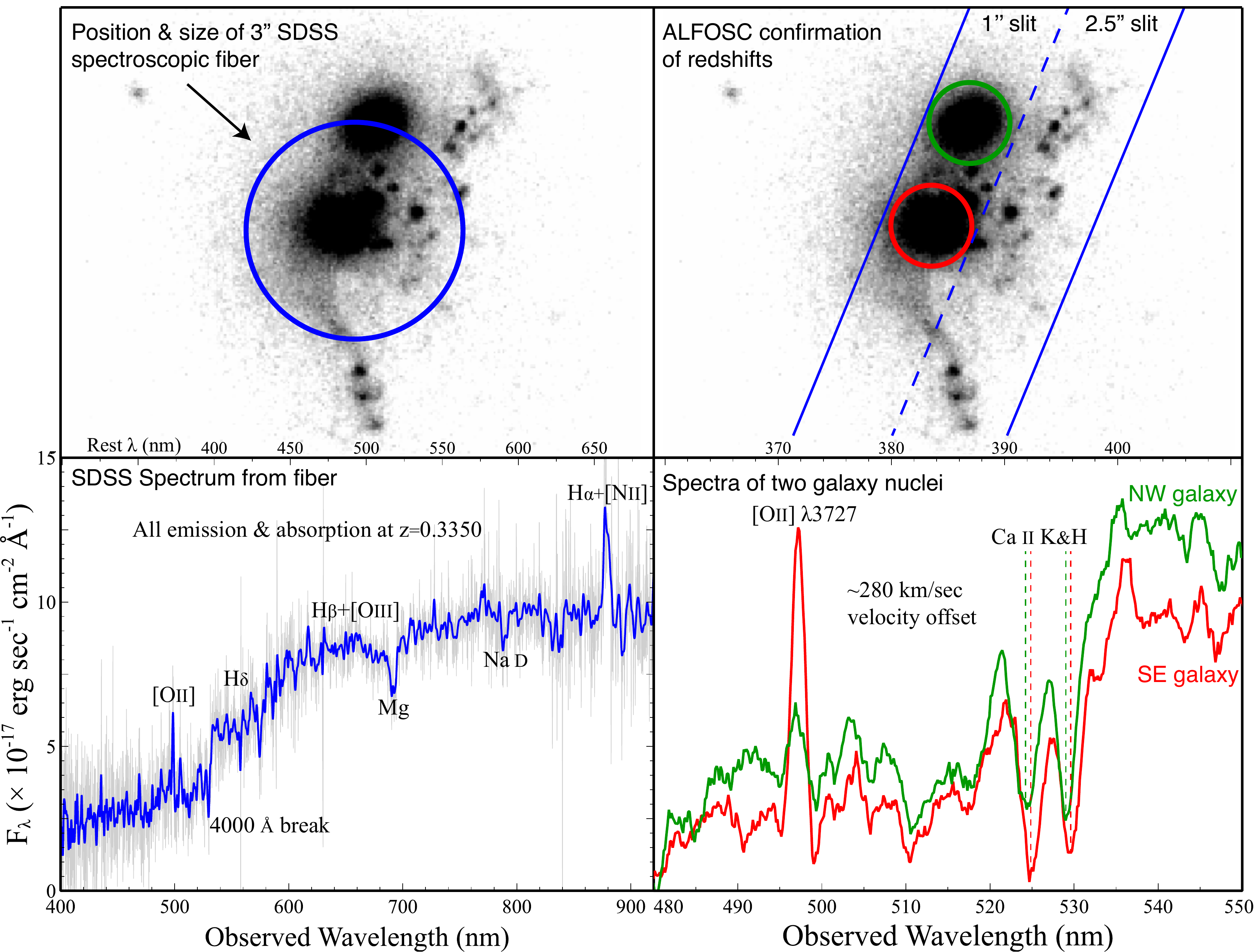}
\end{center}
\vspace*{-3mm}
\caption{({\it left panels}) SDSS
  optical spectrum of the central BCGs, coming from a
  3\arcsec\ diameter spectral fiber whose position and relative size
  is marked by the blue circle on the {\it HST} F606W image at top
  left. All detected emission and absorption lines in the
  spectrum are at a uniform redshift of $z=0.3350\pm0.0002$, including
  H$\alpha$ emission from the young stars and Mg+Na~D absorption and
  4000 \AA\ break from the old stellar population in the
  galaxies. ({\it right panels}) Follow-up NOT/ALFOSC slit spectroscopy. Position and
  P.~A. of the 1\arcsec\ and 2\farcs5\ spectroscopic slits are marked
  on the F606W image in the top right panel. The bottom right panel
  shows extracted spectra from the northwestern (NW) and southeastern
  (SE) halves of the 1\arcsec\ slit in green and red,
  respectively. Emission in both halves of the slit is dominated by
  the two elliptical galaxy nuclei. The galaxies are at the same
  redshift of $z=0.335$, as can be seen by the close wavelength
  alignment of all spectroscopic features. }
\label{fig:fig3}
\end{figure*}

\section{Results}
\subsection{The merging elliptical galaxies}

We present the new {\it HST} data as an RGB composite in Fig.~1. The
images used in this composite are shown individually in Fig.~\ref{fig:fig2}, which
uses contours and labels to highlight various features of interest.
As is evident from these Figures, the cluster center harbors two
elliptical galaxies whose nuclei are separated $\sim
7$ kpc (1\farcs5) in projection from one another. The combined
projected stellar envelope of both galaxies extends $\sim 100$ kpc
($\sim22$\arcsec) across the major axis of the lowest surface
brightness isophote, though their light is very centrally concentrated
with an $r$-band Petrosian radius of $\sim30$ kpc
($\sim7$\arcsec).  Both new and archival optical spectroscopy confirms
that this is no mere projection effect --- the galaxies indeed share a
common redshift, such that their stellar halos must be deeply embedded
in one another over at least $\sim20$ kpc scales.

This is evident in archival Sloan Digital Sky Survey (SDSS,
\citealt{york00}) optical spectroscopy, which we present in the
leftmost panels of Fig.~\ref{fig:fig3}. While the SDSS spectrum is
from a fiber that covers the nuclei of both galaxies (see top left
panel of Fig.~\ref{fig:fig3}), it features a singular line system
including Mg, Na~D, and Ca{\sc~ii h+k} absorption tracing the old
stellar component, all at a common redshift of $z=0.3350\pm0.0002$
(consistent with the SDSS photometric redshift).  This serves as
strong evidence that the galaxies are indeed merging, as they would
have been bright enough to manifest as two distinct-redshift line
systems if indeed they were two otherwise unrelated objects seen in
projection.  

The merger hypothesis is confirmed by follow-up
slit spectroscopy (see Table~\ref{tab:tab1}) obtained with the ALFOSC spectrograph on the Nordic
Optical Telescope (NOT), the results of which are presented in the
rightmost panels of Fig.~\ref{fig:fig3}. As is evident from the
[O{\sc~ii}] and Ca{\sc~ii} lines extracted from the southeastern (SE)
and northwestern (NW) galaxy nuclei (see bottom-right panel), the
maximum spectroscopically permissible velocity offset between the
galaxies is $\sim 280-300$ km sec\mone, far too small for these to be
unrelated galaxies in the Hubble flow. Any 3D configuration consistent
with spectral constraints results in their stellar envelopes being
deeply embedded in one another over $\sim 20$ kpc scales. We therefore
conclude that the two ellipticals (which are of roughly equal stellar
mass, sharing similar surface brightnessess) are undergoing a major
merger.

As noted above, the
cluster mass derived from strong+weak lensing analysis is
$5.13^{+1.33}_{-1.19} \times 10^{14}~h^{-1}$ \Msol\ \citep{oguri12}.  The
implied projected 1D velocity dispersion from such a mass would be of
order $\sim750$ km sec\mone, consistent with the measured velocity
dispersion from 11 member galaxies of $998^{+120}_{-194}$ km sec\mone
\citep{bayliss11}. The expectation value for pairwise
velocities along the line of sight would therefore be $\gtrsim 1000$ km sec\mone,
rather more than the $\sim 280$ km sec\mone permitted by our
follow-up spectroscopy.  
The NOT data therefore suggest that the trajectories of the merger
participants lie mostly in the plane of the sky.  
While the galaxy stellar isophotes are mostly
smooth, unsharp masks and a F606W/F160W color map reveal residual
surface brightness fluctuations between and around both nuclei,
suggestive of some dynamical disturbance in the old stellar component
as a result of the merger.

\begin{figure}
\begin{center}
\includegraphics[scale=0.7]{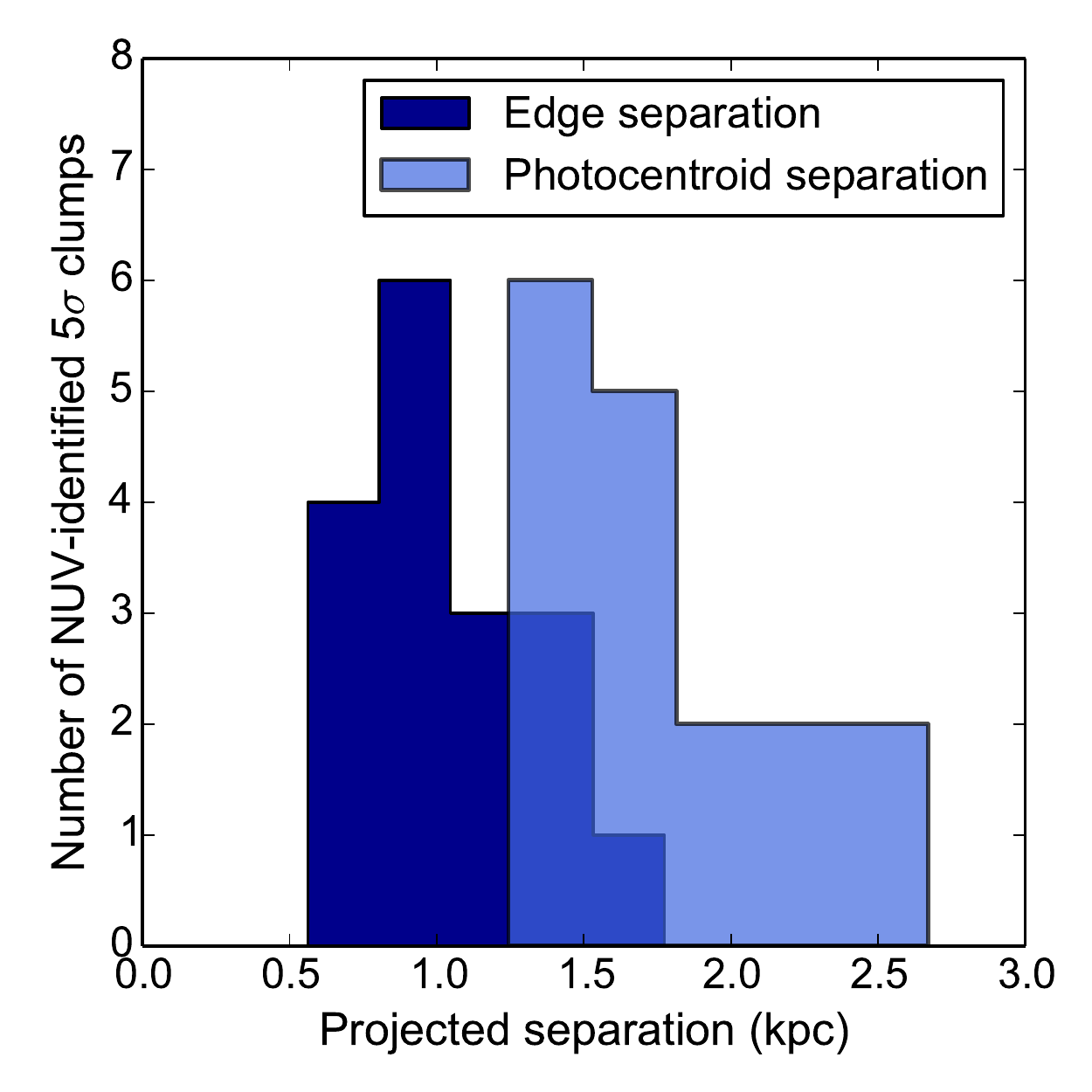}      
\end{center}
\vspace*{-6mm}
\caption{Projected separations, in kpc, between adjacent NUV-bright clumps associated 
with $\gae5\sigma$ count rate overdensities in the {\it HST} F390W image. The light blue histogram 
shows measured separations between clump centroids, and the dark blue histogram shows 
separation between the {\it edges} of those adjacent clumps, where a clump edge is roughly defined as the 
radial boundary encompassing $\sim90\%$ of all flux associated with the clump. 
Clump centroids are generally spaced 
$\sim 1.5$ kpc in projection from one another, whereas the projected separation between the edges of clumps is $\sim 1$ kpc. 
Typical resolved 
clump diameters are $\sim 1$ kpc. On galaxy-wide scales, the ``beads on a string'' mode of star 
formation is believed to be associated with 
the formation of Jeans length superclusters separated by $1-2$ kpc from one another, consistent with what is observed here. 
}  
\label{fig:histogram}
\end{figure}

\subsection{The young super star clusters}

Far  more dramatic than the merging galaxies  is the  F390W  image of  rest-frame NUV  continuum
emission mostly arising  from young ($\lae 300$ Myr),  massive ($\gae 5$
\Msol) stars. This  is seen in blue on the  rightmost panel of Fig.~1,
and shown in the bottom-left panel  of Fig.~2. 
The NUV continuum emission is shown in green contours on the F160W image
in the top-left panel of  Fig.~2. 
Not counting the  nuclei of the two elliptical  galaxies, the emission
consists  of at least 19 stellar  superclusters (or perhaps tidal dwarf galaxies) 
$\sim0.5-1$  kpc in
diameter. The 19 stellar clusters are  enhanced in NUV surface  brightness by
factors  of $2-10$  over  the  filaments in which they are embedded.  
Each  of  these superclusters  are
numerically labeled  in the bottom-right panel of  Fig.~2, which shows
an unsharp  mask revealing residual  [O~{\sc ii}] and blue excess emission in  the F606W
image.

The
young  stellar superclusters  are roughly  equally spaced  along $3-4$
narrow ($1-2$ kpc  wide), straight filaments roughly $\sim  10$ kpc in
projected length, resembling ``beads on a string'' (discussed in \S4).
As we show in Fig.~\ref{fig:histogram}, a histogram  of projected  separations  between neighboring stellar clusters
reveals a  strong peak around  $\sim 1-2$ kpc. 
All 19 clusters are at least marginally resolved, and  while most are
roughly circular,  a few (especially cluster $\#$2)  show evidence for
$\sim500$ pc-scale asymmetries reminiscent of tidal arms.

Two filamentary strings of clusters (containing clusters numbered
$6-9$ and $15-19$ on the bottom-right panel of Fig.~2) are aligned
along a position angle that is roughly parallel to the axis joining
the nuclei of the two merging galaxies (P.~A.$\sim -45^\circ$,
measured North through East).  The other two strings (containing
clusters $1-4$ and $11-14$, excluding \#13) lie along a P.~A.$\approx 45^\circ$ that
is roughly perpendicular to this axis.  The clusters are entirely
contained within the combined projected old stellar envelope of both
galaxies, within a circle of diameter $\sim 27$ kpc (5\farcs6). This
can be seen in the top-left panel of Fig.~2, in which the NUV
continuum emission is plotted in green contours over the F160W image,
showing the old stellar population.

\section{Discussion}

\subsection{Origin of the ultraviolet emission}

We consider three scenarios for the physical nature and origin of the clumpy NUV continuum 
emission. 

\begin{enumerate}

\item It is a gravitationally lensed image (or a set of images) from a
  higher redshift background galaxy (or multiple galaxies).

\item It is a chance  superposition along the line of sight, arising
  instead from a projection effect due to unrelated foreground
  sources.

\item It arises from ongoing or very recent star formation taking
  place within the stellar envelope of the interacting elliptical
  galaxies.

\end{enumerate}

Scenarios (1) and (2) are entirely ruled out by spectroscopic
evidence.  The same SDSS and NOT spectroscopy used to confirm that the
two elliptical galaxies are indeed merging also confirms that
H$\alpha$ and [O~{\sc ii}] emission from star formation is mapped to
the same redshift of the galaxies at $z=0.3350\pm0.0002$ (see
Fig.~\ref{fig:fig3}). The stellar superclusters therefore must inhabit
the combined stellar envelope of the merger participants.  The
integrated flux density of the NUV continuum in the F390W bandpass is
$\left(5.161 \pm 0.512\right) \times 10^{-17}$ erg
sec\mone\ cm\mtwo\ \AA\mone, suggesting that line emission from the
young stellar superclusters is easily bright enough to be detected in
the spectroscopy .  If at appreciably different redshift than the two
galaxies with which the emission is cospatial, the redshift of the
Balmer lines should be offset from that of the Mg and Na~D absorption
as well as the calcium break.  
Moreover, strong lensing analysis (Sharon et al.~2014, submitted to ApJ) indicates that possible lensed images of background sources do not contribute significantly to the flux seen in the NUV emission.
For the remainder of this
Letter, we therefore adopt the most evidence-supported scenario (3),
and conclude that (a) the two elliptical galaxies seen in close
separation are indeed undergoing a major merger and (b) the clumpy NUV
continuum emission arises from ongoing or recent star formation taking
place within their combined stellar envelope.

\subsection{Properties of the young stellar superclusters}

The integrated SDSS H$\alpha$ line flux from the 3\arcsec\ diameter
fiber, which covers only the innermost region of the star forming
chain (see Fig.~\ref{fig:fig3}, top right), is $\left(6.335\pm1.77\right) \times 10^{-16}$ erg
sec\mone\ cm\mtwo\ with a FWHM of $152\pm15$ km sec\mone.
We estimate internal extinction via the Balmer decrement
(H$\alpha$/H$\beta$ flux ratio), following the procedure described in
\citet{tremblay10} and adopting ``case B'' recombination and the $R_V=3.1$ reddening-to-extinction
law of \citet{cardelli89}.  After an additional correction for
foreground Milky Way extinction using $E(B-V)=0.023$, we estimate an
extinction-corrected H$\alpha$ luminosity of $\left(3.65\pm1.1\right) \times 10^{41}$
erg sec\mone\ from the region covered by the SDSS fiber.
As part of our NOT/ALFOSC observations, we used a 2\farcs5 wide
slit to observe the stellar superclusters in H$\alpha$, as
detailed in Figure 3. This slit encompasses all of the NUV identified
clumps, even after accounting for minimal slit losses given ground-based seeing. The derived H$\alpha$ line flux from our ALFOSC data is 1.7 times greater than
that reported by SDSS, approximately as expected given the more complete
spatial coverage of the clumps by the wide slit observation. Including a 30\% uncertainty to account for spectrograph CCD fringing and our [N{\sc~ii}] contamination correction, the
implied total extinction-corrected H$\alpha$ 
luminosity for the entire system is 
$\left(6.21 \pm 1.9 \right)\times 10^{41}$ erg sec\mone.

By Eq.~2 in \citet{kennicutt98}, this extinction-corrected H$\alpha$ luminosity
corresponds to a (very rough and assumption-heavy) star formation rate
(SFR) of $\sim5\pm2$ \Msol\ yr\mone.  Automated
modeling of the SDSS spectrophotometric data by \citet{maraston09,maraston06,maraston13}, 
which matches SED templates to 
extinction-corrected SDSS $ugriz$ magnitudes, 
yields an SFR of $9.55\pm3.4$ \Msol\ yr\mone\ and a total stellar mass
of $\left(3.35\pm0.04\right)\times10^{11}$ \Msol\ (for more details, see \citealt{maraston06}).  Absent the
follow-up data needed for a more robust analysis, 
the remainder of this Letter assumes that the SFR is roughly 
in the range of $\sim 5-10$ \Msol\ yr\mone.  Each
clump contributes approximately $\sim 2-6\%$ of the total NUV flux,
which should roughly scale to its relative contribution to the total
star formation rate. We therefore estimate that the clumps have an
associated star formation rate in the range of $\sim 0.1-0.6$
\Msol\ yr\mone.  

Inverting the \citet{bigiel08} calibration of the molecular Schmidt law \citep{kennicutt98} and assuming gas depletion times 
in the range of $1-2$ Gyr,  the total SFR translates to a rough associated
molecular gas mass of $M_{\mathrm{H}_2} \approx 0.5-2 \times10^{10}$
\Msol. We obtained an IRAM 30m telescope CO$(1-0)$ observation 
in an attempt to measure the actual cold gas mass. 
No line was detected at a sensitivity of 0.9 mK $T_{\mathrm{mb}}$ per 80 km sec\mone\ channel, setting a 1$\sigma$ upper limit of $\sim1\times10^{10}$ \Msol\  of molecular gas, 
assuming a line width of 220 km s\mone\ 
and a Galactic $X_{\mathrm{CO}}$ factor.
A molecular gas mass in the range of  $\sim 0.5-1 \times10^{10}$ \Msol would be $\sim 1-3\%$ that of the galaxies'  
stellar mass, consistent with that observed in large samples of early type 
galaxies (e.g., \citealt{young11}).

The observed star formation in SDSS 1531 is strongly reminiscent of
the well-known ``beads on a string'' mode of star formation (e.g.,
\citealt{chandra53,toomre77,elmegreen96,renaud08}) frequently observed
in the arms of spiral galaxies \citep{elmegreen83}, resonance rings
\citep{elmegreen94}, and tidal arms that bridge interacting galaxies
\citep{duc00,smith10}.  Beads on a string star formation is a
kpc-scale manifestation of the Jeans length, and the physics governing
the system is analogous to the Plateau-Rayleigh instability causing
(e.g.) a continuous column of falling water to disrupt, explaining why
rain falls in drops rather than in unbroken filaments from the sky
(e.g., \citealt{quillen10}).  
Despite being broadly categorized as ``red and dead'', it has been known 
for a number of years that star formation can be relatively abundant in both cluster
and field ellipticals \citep{yi05,odea08,jeong09,davis11,tremblay12a}.
Nevertheless, the formation of ``beads on a string'' stellar superclusters in major, likely gas-rich mergers between giant ellipticals is rarely (if ever) observed, 
regardless of whether or not the star formation is driven by a merger 
or a cooling flow (e.g., Tremblay et al.~2014b, in prep).
The unusual system in
SDSS 1531 is likely a serendipitous snapshot of what may be a
very short-lived morphology, presenting a unique opportunity to study
star formation and gas dynamical response in a tidal field governed by
dynamical friction, shear, and gravitational torques associated with
the two merging ellipticals. Forthcoming cluster mass reconstruction
from strong lensing (Sharon et al.~2014, submitted to ApJ.) will provide a
well-constrained canvas against which these kinematics may be studied.

These quantitative analyses, however, await the necessary
multiwavelength follow-up data. Recently obtained
Gemini GMOS-N IFU observations (Tremblay et al.~2014c, in prep) will
disentangle the kinematics, dynamical timescales, and 3D geometry of
the system, as well as local internal extinction corrections required
for a better SFR estimate.  ALMA is the only
facility with the sensitivity and resolution needed to independently
constrain the clump-by-clump SFR and molecular gas masses.

 Finally, {\it Chandra} X-ray observations
would determine whether or not the observed star formation might
originate from gas that has condensed from the ambient hot atmosphere
via a cooling flow, or a shock in the X-ray gas driven by colliding
hot halos from the galaxy merger. The location of the stellar
superclusters leaves this origin question open, as the star formation
is not occurring in a region between the merging galaxies, where one
would most obviously expect it if the star formation is shock
triggered. However, viscous drag effects preferentially influencing
the gas may be responsible for the apparent dislocation of the stellar
(and presumably dark matter mass) components from the gas, as has been
observed in other galaxy cluster mergers (e.g., \citealt{clowe06}).\\

\acknowledgments We thank Profs.~Fran\c{c}oise Combes, Eric Emsellem,
and Tim de Zeeuw for thoughtful discussions. We also thank 
the anonymous referee whose feedback improved this work. G.R.T. and T.A.D.~acknowledge
support from a European Southern Observatory (ESO) Fellowship
partially funded by the European Community's Seventh Framework
Programme (/FP7/2007-2013/) under grant agreement No.~229517.  
Support for program number HST-GO-13003 was provided by NASA through a grant from the Space Telescope Science Institute, which is operated by the Association of Universities for Research in Astronomy, Inc., under NASA contract NAS5-26555. 
This
paper is based on observations by the NASA/ESA {\it Hubble Space
  Telescope}, obtained at the Space Telescope Science Institute.
We also present results from the 
the Nordic Optical
  Telescope, operated by the Nordic Optical Telescope Scientific Association at the 
Observatorio del Roque de los Muchachos, La Palma, Spain, of the 
Instituto de Astrofisica de Canarias.  
Funding for the SDSS
and SDSS-II has been provided by the Alfred P.~Sloan Foundation, the
Participating Institutions, the National Science Foundation, the
U.S.~Department of Energy, the National Aeronautics and Space
Administration, the Japanese Monbukagakusho, the Max Planck Society,
and the Higher Education Funding Council for England. The SDSS Web
Site is \url{http://www.sdss.org/}.


\begin{thebibliography}{}
\expandafter\ifx\csname natexlab\endcsname\relax\def\natexlab#1{#1}\fi

\bibitem[{{Bayliss} {et~al.}(2011){Bayliss}, {Hennawi}, {Gladders}, {Koester},
  {Sharon}, {Dahle}, \& {Oguri}}]{bayliss11}
{Bayliss}, M.~B., {Hennawi}, J.~F., {Gladders}, M.~D., {et~al.} 2011, \apjs,
  193, 8

\bibitem[{{Bayliss} {et~al.}(2013){Bayliss}, {Rigby}, {Sharon}, {Wuyts},
  {Florian}, {Gladders}, {Johnson}, \& {Oguri}}]{bayliss13}
{Bayliss}, M.~B., {Rigby}, J.~R., {Sharon}, K., {et~al.} 2013, ArXiv e-prints,
  arXiv:1310.6695

\bibitem[{{Bigiel} {et~al.}(2008){Bigiel}, {Leroy}, {Walter}, {Brinks}, {de
  Blok}, {Madore}, \& {Thornley}}]{bigiel08}
{Bigiel}, F., {Leroy}, A., {Walter}, F., {et~al.} 2008, \aj, 136, 2846

\bibitem[{{Cardelli} {et~al.}(1989){Cardelli}, {Clayton}, \&
  {Mathis}}]{cardelli89}
{Cardelli}, J.~A., {Clayton}, G.~C., \& {Mathis}, J.~S. 1989, \apj, 345, 245

\bibitem[{{Chandrasekhar} \& {Fermi}(1953)}]{chandra53}
{Chandrasekhar}, S., \& {Fermi}, E. 1953, \apj, 118, 116

\bibitem[{{Clowe} {et~al.}(2006){Clowe}, {Schneider}, {Arag{\'o}n-Salamanca},
  {Bremer}, {De Lucia}, {Halliday}, {Jablonka}, {Milvang-Jensen}, {Pell{\'o}},
  {Poggianti}, {Rudnick}, {Saglia}, {Simard}, {White}, \& {Zaritsky}}]{clowe06}
{Clowe}, D., {Schneider}, P., {Arag{\'o}n-Salamanca}, A., {et~al.} 2006, \aap,
  451, 395

\bibitem[{{Coe} {et~al.}(2012){Coe}, {Umetsu}, {Zitrin}, {Donahue},
  {Medezinski}, {Postman}, {Carrasco}, {Anguita}, {Geller}, {Rines},
  {Diaferio}, {Kurtz}, {Bradley}, {Koekemoer}, {Zheng}, {Nonino}, {Molino},
  {Mahdavi}, {Lemze}, {Infante}, {Ogaz}, {Melchior}, {Host}, {Ford}, {Grillo},
  {Rosati}, {Jim{\'e}nez-Teja}, {Moustakas}, {Broadhurst}, {Ascaso}, {Lahav},
  {Bartelmann}, {Ben{\'{\i}}tez}, {Bouwens}, {Graur}, {Graves}, {Jha},
  {Jouvel}, {Kelson}, {Moustakas}, {Maoz}, {Meneghetti}, {Merten}, {Riess},
  {Rodney}, \& {Seitz}}]{coe12}
{Coe}, D., {Umetsu}, K., {Zitrin}, A., {et~al.} 2012, \apj, 757, 22

\bibitem[{{Davis} {et~al.}(2011){Davis}, {Alatalo}, {Sarzi}, {Bureau}, {Young},
  {Blitz}, {Serra}, {Crocker}, {Krajnovi{\'c}}, {McDermid}, {Bois}, {Bournaud},
  {Cappellari}, {Davies}, {Duc}, {de Zeeuw}, {Emsellem}, {Khochfar},
  {Kuntschner}, {Lablanche}, {Morganti}, {Naab}, {Oosterloo}, {Scott}, \&
  {Weijmans}}]{davis11}
{Davis}, T.~A., {Alatalo}, K., {Sarzi}, M., {et~al.} 2011, \mnras, 417, 882

\bibitem[{{Dressel}(2012)}]{dressel12}
{Dressel}, L. 2012, {Wide Field Camera 3 Instrument Handbook for Cycle 21 v.
  5.0}

\bibitem[{{Duc} {et~al.}(2000){Duc}, {Brinks}, {Springel}, {Pichardo},
  {Weilbacher}, \& {Mirabel}}]{duc00}
{Duc}, P.-A., {Brinks}, E., {Springel}, V., {et~al.} 2000, \aj, 120, 1238

\bibitem[{{Elmegreen}(1994)}]{elmegreen94}
{Elmegreen}, B.~G. 1994, \apjl, 425, L73

\bibitem[{{Elmegreen} \& {Efremov}(1996)}]{elmegreen96}
{Elmegreen}, B.~G., \& {Efremov}, Y.~N. 1996, \apj, 466, 802

\bibitem[{{Elmegreen} \& {Elmegreen}(1983)}]{elmegreen83}
{Elmegreen}, B.~G., \& {Elmegreen}, D.~M. 1983, \mnras, 203, 31

\bibitem[{{Fruchter et al.}(2010)}]{fruchter10}
{Fruchter et al.} 2010, in 2010 Space Telescope Science Institute Calibration
  Workshop, p. 382-387, 382--387

\bibitem[{{Gralla} {et~al.}(2011){Gralla}, {Sharon}, {Gladders}, {Marrone},
  {Barrientos}, {Bayliss}, {Bonamente}, {Bulbul}, {Carlstrom}, {Culverhouse},
  {Gilbank}, {Greer}, {Hasler}, {Hawkins}, {Hennessy}, {Joy}, {Koester},
  {Lamb}, {Leitch}, {Miller}, {Mroczkowski}, {Muchovej}, {Oguri}, {Plagge},
  {Pryke}, \& {Woody}}]{gralla11}
{Gralla}, M.~B., {Sharon}, K., {Gladders}, M.~D., {et~al.} 2011, \apj, 737, 74

\bibitem[{{Hennawi} {et~al.}(2008){Hennawi}, {Gladders}, {Oguri}, {Dalal},
  {Koester}, {Natarajan}, {Strauss}, {Inada}, {Kayo}, {Lin}, {Lampeitl},
  {Annis}, {Bahcall}, \& {Schneider}}]{hennawi08}
{Hennawi}, J.~F., {Gladders}, M.~D., {Oguri}, M., {et~al.} 2008, \aj, 135, 664

\bibitem[{{Jeong} {et~al.}(2009){Jeong}, {Yi}, {Bureau}, {Davies},
  {Falc{\'o}n-Barroso}, {van de Ven}, {Peletier}, {Bacon}, {Cappellari}, {de
  Zeeuw}, {Emsellem}, {Krajnovi{\'c}}, {Kuntschner}, {McDermid}, {Sarzi}, \&
  {van den Bosch}}]{jeong09}
{Jeong}, H., {Yi}, S.~K., {Bureau}, M., {et~al.} 2009, \mnras, 398, 2028

\bibitem[{{Kaviraj} {et~al.}(2012){Kaviraj}, {Darg}, {Lintott}, {Schawinski},
  \& {Silk}}]{kaviraj12}
{Kaviraj}, S., {Darg}, D., {Lintott}, C., {Schawinski}, K., \& {Silk}, J. 2012,
  \mnras, 419, 70

\bibitem[{{Kennicutt}(1998)}]{kennicutt98}
{Kennicutt}, Jr., R.~C. 1998, \apj, 498, 541

\bibitem[{{Maraston} {et~al.}(2006){Maraston}, {Daddi}, {Renzini}, {Cimatti},
  {Dickinson}, {Papovich}, {Pasquali}, \& {Pirzkal}}]{maraston06}
{Maraston}, C., {Daddi}, E., {Renzini}, A., {et~al.} 2006, \apj, 652, 85

\bibitem[{{Maraston} {et~al.}(2009){Maraston}, {Str{\"o}mb{\"a}ck}, {Thomas},
  {Wake}, \& {Nichol}}]{maraston09}
{Maraston}, C., {Str{\"o}mb{\"a}ck}, G., {Thomas}, D., {Wake}, D.~A., \&
  {Nichol}, R.~C. 2009, \mnras, 394, L107

\bibitem[{{Maraston} {et~al.}(2013){Maraston}, {Pforr}, {Henriques}, {Thomas},
  {Wake}, {Brownstein}, {Capozzi}, {Tinker}, {Bundy}, {Skibba}, {Beifiori},
  {Nichol}, {Edmondson}, {Schneider}, {Chen}, {Masters}, {Steele}, {Bolton},
  {York}, {Weaver}, {Higgs}, {Bizyaev}, {Brewington}, {Malanushenko},
  {Malanushenko}, {Snedden}, {Oravetz}, {Pan}, {Shelden}, \&
  {Simmons}}]{maraston13}
{Maraston}, C., {Pforr}, J., {Henriques}, B.~M., {et~al.} 2013, \mnras, 435,
  2764

\bibitem[{{O'Dea} {et~al.}(2008){O'Dea}, {Baum}, {Privon}, {Noel-Storr},
  {Quillen}, {Zufelt}, {Park}, {Edge}, {Russell}, {Fabian}, {Donahue},
  {Sarazin}, {McNamara}, {Bregman}, \& {Egami}}]{odea08}
{O'Dea}, C.~P., {Baum}, S.~A., {Privon}, G., {et~al.} 2008, \apj, 681, 1035

\bibitem[{{Oguri} {et~al.}(2012){Oguri}, {Bayliss}, {Dahle}, {Sharon},
  {Gladders}, {Natarajan}, {Hennawi}, \& {Koester}}]{oguri12}
{Oguri}, M., {Bayliss}, M.~B., {Dahle}, H., {et~al.} 2012, \mnras, 420, 3213

\bibitem[{{Oguri} {et~al.}(2009){Oguri}, {Hennawi}, {Gladders}, {Dahle},
  {Natarajan}, {Dalal}, {Koester}, {Sharon}, \& {Bayliss}}]{oguri09}
{Oguri}, M., {Hennawi}, J.~F., {Gladders}, M.~D., {et~al.} 2009, \apj, 699,
  1038

\bibitem[{{Postman} {et~al.}(2012){Postman}, {Coe}, {Ben{\'{\i}}tez},
  {Bradley}, {Broadhurst}, {Donahue}, {Ford}, {Graur}, {Graves}, {Jouvel},
  {Koekemoer}, {Lemze}, {Medezinski}, {Molino}, {Moustakas}, {Ogaz}, {Riess},
  {Rodney}, {Rosati}, {Umetsu}, {Zheng}, {Zitrin}, {Bartelmann}, {Bouwens},
  {Czakon}, {Golwala}, {Host}, {Infante}, {Jha}, {Jimenez-Teja}, {Kelson},
  {Lahav}, {Lazkoz}, {Maoz}, {McCully}, {Melchior}, {Meneghetti}, {Merten},
  {Moustakas}, {Nonino}, {Patel}, {Reg{\"o}s}, {Sayers}, {Seitz}, \& {Van der
  Wel}}]{postman12}
{Postman}, M., {Coe}, D., {Ben{\'{\i}}tez}, N., {et~al.} 2012, \apjs, 199, 25

\bibitem[{{Quillen} \& {Comparetta}(2010)}]{quillen10}
{Quillen}, A.~C., \& {Comparetta}, J.\ 2010, ArXiv e-prints, arXiv:1002.4870


\bibitem[{{Renaud} {et~al.}(2008){Renaud}, {Boily}, {Fleck}, {Naab}, \&
  {Theis}}]{renaud08}
{Renaud}, F., {Boily}, C.~M., {Fleck}, J.-J., {Naab}, T., \& {Theis}, C. 2008,
  \mnras, 391, L98

\bibitem[{{Smith} {et~al.}(2010){Smith}, {Khosroshahi}, {Dariush}, {Sanderson},
  {Ponman}, {Stott}, {Haines}, {Egami}, \& {Stark}}]{smith10}
{Smith}, G.~P., {Khosroshahi}, H.~G., {Dariush}, A., {et~al.} 2010, \mnras,
  409, 169

\bibitem[{{Toomre}(1977)}]{toomre77}
{Toomre}, A. 1977, in Evolution of Galaxies and Stellar Populations, ed. B.~M.
  {Tinsley} \& R.~B.~G. {Larson}, D.~Campbell, 401

\bibitem[{{Tremblay} {et~al.}(2010){Tremblay}, {O'Dea}, {Baum}, {Koekemoer},
  {Sparks}, {de Bruyn}, \& {Schoenmakers}}]{tremblay10}
{Tremblay}, G.~R., {O'Dea}, C.~P., {Baum}, S.~A., {et~al.} 2010, \apj, 715, 172

\bibitem[{{Tremblay} {et~al.}(2012){Tremblay}, {O'Dea}, {Baum}, {Clarke},
  {Sarazin}, {Bregman}, {Combes}, {Donahue}, {Edge}, {Fabian}, {Ferland},
  {McNamara}, {Mittal}, {Oonk}, {Quillen}, {Russell}, {Sanders}, {Salom{\'e}},
  {Voit}, {Wilman}, \& {Wise}}]{tremblay12a}
---. 2012, \mnras, 424, 1042

\bibitem[{{Yi} {et~al.}(2005){Yi}, {Yoon}, {Kaviraj}, {Deharveng}, {Rich},
  {Salim}, {Boselli}, {Lee}, {Ree}, {Sohn}, {Rey}, {Lee}, {Rhee}, {Bianchi},
  {Byun}, {Donas}, {Friedman}, {Heckman}, {Jelinsky}, {Madore}, {Malina},
  {Martin}, {Milliard}, {Morrissey}, {Neff}, {Schiminovich}, {Siegmund},
  {Small}, {Szalay}, {Jee}, {Kim}, {Barlow}, {Forster}, {Welsh}, \&
  {Wyder}}]{yi05}
{Yi}, S.~K., {Yoon}, S.-J., {Kaviraj}, S., {et~al.} 2005, \apjl, 619, L111

\bibitem[{{York} {et~al.}(2000){York}, {Adelman}, {Anderson}, {Anderson},
  {Annis}, {Bahcall}, {Bakken}, {Barkhouser}, {Bastian}, {Berman}, {Boroski},
  {Bracker}, {Briegel}, {Briggs}, {Brinkmann}, {Brunner}, {Burles}, {Carey},
  {Carr}, {Castander}, {Chen}, {Colestock}, {Connolly}, {Crocker}, {Csabai},
  {Czarapata}, {Davis}, {Doi}, {Dombeck}, {Eisenstein}, {Ellman}, {Elms},
  {Evans}, {Fan}, {Federwitz}, {Fiscelli}, {Friedman}, {Frieman}, {Fukugita},
  {Gillespie}, {Gunn}, {Gurbani}, {de Haas}, {Haldeman}, {Harris}, {Hayes},
  {Heckman}, {Hennessy}, {Hindsley}, {Holm}, {Holmgren}, {Huang}, {Hull},
  {Husby}, {Ichikawa}, {Ichikawa}, {Ivezi{\'c}}, {Kent}, {Kim}, {Kinney},
  {Klaene}, {Kleinman}, {Kleinman}, {Knapp}, {Korienek}, {Kron}, {Kunszt},
  {Lamb}, {Lee}, {Leger}, {Limmongkol}, {Lindenmeyer}, {Long}, {Loomis},
  {Loveday}, {Lucinio}, {Lupton}, {MacKinnon}, {Mannery}, {Mantsch}, {Margon},
  {McGehee}, {McKay}, {Meiksin}, {Merelli}, {Monet}, {Munn}, {Narayanan},
  {Nash}, {Neilsen}, {Neswold}, {Newberg}, {Nichol}, {Nicinski}, {Nonino},
  {Okada}, {Okamura}, {Ostriker}, {Owen}, {Pauls}, {Peoples}, {Peterson},
  {Petravick}, {Pier}, {Pope}, {Pordes}, {Prosapio}, {Rechenmacher}, {Quinn},
  {Richards}, {Richmond}, {Rivetta}, {Rockosi}, {Ruthmansdorfer}, {Sandford},
  {Schlegel}, {Schneider}, {Sekiguchi}, {Sergey}, {Shimasaku}, {Siegmund},
  {Smee}, {Smith}, {Snedden}, {Stone}, {Stoughton}, {Strauss}, {Stubbs},
  {SubbaRao}, {Szalay}, {Szapudi}, {Szokoly}, {Thakar}, {Tremonti}, {Tucker},
  {Uomoto}, {Vanden Berk}, {Vogeley}, {Waddell}, {Wang}, {Watanabe},
  {Weinberg}, {Yanny}, {Yasuda}, \& {SDSS Collaboration}}]{york00}
{York}, D.~G., {Adelman}, J., {Anderson}, Jr., J.~E., {et~al.} 2000, \aj, 120,
  1579

\bibitem[{{Young} {et~al.}(2011){Young}, {Bureau}, {Davis}, {Combes},
  {McDermid}, {Alatalo}, {Blitz}, {Bois}, {Bournaud}, {Cappellari}, {Davies},
  {de Zeeuw}, {Emsellem}, {Khochfar}, {Krajnovi{\'c}}, {Kuntschner},
  {Lablanche}, {Morganti}, {Naab}, {Oosterloo}, {Sarzi}, {Scott}, {Serra}, \&
  {Weijmans}}]{young11}
{Young}, L.~M., {Bureau}, M., {Davis}, T.~A., {et~al.} 2011, \mnras, 414, 940

\end{thebibliography}
\end{document}